\documentclass[aps,prl,showpacs,twocolumn,amsmath,amssymb,superscriptaddress,footinbib]{revtex4}

\usepackage[english]{babel}
\usepackage{latexsym}
\usepackage{graphicx}
\usepackage{subfigure}
\usepackage{epsfig}
\usepackage{amsfonts}
\usepackage{amssymb}
\usepackage{amsmath}
\usepackage{bbm}

\begin{document}

\title{Absence of Vacuum Induced Berry Phases without the Rotating Wave Approximation in Cavity QED}

%%%%%%%%%%%%%%%%%%%%%%%%%%%%%%%%%%%%%%%%%%%%%%%%%%%%%%%%%%%%%%%%%%%%%%%%%%%%%%%%%%

\author{Jonas Larson}
\email{jolarson@physto.su.se} \affiliation{Department of Physics,
Stockholm University, AlbaNova University Center, Se-106 91 Stockholm,
Sweden}
\affiliation{Institut f\"ur Theoretische Physik, Universit\"at zu K\"oln, K\"oln, De-50937, Germany}

\date{\today}

\begin{abstract}
We revisit earlier studies on Berry phases suggested to appear in certain cavity QED settings. It has been especially argued that a non-trivial geometric phase is achievable even in the situation of no cavity photons. We, however, show that such results hinge on imposing the {\it rotating wave approximation} (RWA), while without the RWA no Berry phases occur in these schemes. A geometrical interpretation of our results is obtained by introducing semi-classical energy surfaces which in a simple way brings out the phase space dynamics. With the RWA, a conical intersection between the surfaces emerges and encircling it gives rise to the Berry phase. Without the RWA, the conical intersection is absent and therefore the Berry phase vanishes. It is believed that this is a first example showing how the application of the RWA in the Jaynes-Cummings model may lead to false conclusions, regardless mutual strengths between the system parameters.
\end{abstract}

\pacs{45.50.Pq, 03.65.Vf, 42.50.Ct}
\maketitle

The Jaynes-Cummings (JC) model~\cite{jc} has successfully served as a work horse in cavity electrodynamics (QED) for more than three decades. Despite being extremely simple and analytically solvable it manages to theoretically explain many of the QED experiments to date, both in the microwave~\cite{micro} and optical~\cite{optical} regimes, as well as more recent experiments on superconducting qubits coupled to transmission line resonators~\cite{circuit}. There are, however, exceptions where the underlying approximations of the JC model break down and the model render erroneous predictions. A known and controversial example is the rule out of the so called Dicke phase transition~\cite{dickept} due to ignoring the self-energy of the electromagnetic field~\cite{nogo}. In this Letter we give another example where instead application of the RWA yield wrong results concerning expected Berry phases. 

The first examples of quantum geometric phases date back to the late 1950s in works by Aharonov and Bohm~\cite{ab} and Longuet-Higgins {\it et al.}~\cite{mol}. However, it was not until the seminal paper by Berry in 1984~\cite{berry}, where a more general description, and thereby also deeper understanding, of these phases were outlined, that interest in the topic seriously took off~\cite{berrybook}. Over the years, various extensions of the phase have been considered, as for example in Ref.~\cite{berryext}. Nowadays it is known that the Berry phase is not only of interest from a fundamental point of view, for example, it turns out to be deeply rooted with topological states of matter~\cite{topins}, and it may turn out to have important applications in quantum computing~\cite{quantinf}. 

In 2002, I. Fuentes-Guridi {\it et al.} modified the archetype of Berry phases, namely a spin-1/2 particle in a classical magnetic field~\cite{vedral1}. The external classical field was replaced by a fully quantum one, and the Berry phase was analyzed in terms of the above mentioned JC model. Such extension is most interesting since; ($i$) the spin-1/2 particle and the field form a composite quantum system where the driving field must be treated on the same footing as the particle itself, and ($ii$) in the limit of zero photons it is not {\it a priori} known whether quantum vacuum fluctuations can give rise to a Berry phase. Clearly, such vacuum induced Berry phase would have no counterpart in a model constructed by a classical driving field, hence it would be a direct proof of electromagnetic field quantization. The novel work of I. Fuentes-Guridi {\it et al.} avalanched a series of following papers~\cite{ext1,ext2,dickeberry,gate,deco,qubit,multiatom}. In particular, various extensions of Ref.~\cite{vedral1} have been addressed, such as; multi-atom (Dicke) systems~\cite{dickeberry}, geometrical quantum computing~\cite{gate}, decohering cavity modes~\cite{deco}, solid state counterparts~\cite{qubit}, and multi-level atom situations~\cite{multiatom}. Most of these references picture one or another sort of cavity QED scheme, and more importantly the RWA has been imposed in all of them. The main purpose of the present Letter is to demonstrate that the presence of a non-trivial Berry phase is indeed an outcome of applying the RWA. For this aim we construct energy surfaces where the contour-lines represents semi-classical phase space trajectories. In the JC model, these trajectories encircle a conical intersection (CI) which is the origin of the non-vanishing Berry phase. Without the RWA, on the other hand, there exist no CI and as a consequence the Berry phase is strictly zero in this case. These semi-classical conclusions are supported by direct numerical diagonalization of the full quantum system. Our findings are particularly interesting since contrary to earlier discussions on the RWA in the JC model, the flaw in its applicability discussed here is independent on the strength of the system parameters. The origin in such surprising results lies in the adiabatic assumption, in this limit the {\it counter-rotating terms} related to the RWA cannot be omitted.

We begin by reviewing the idea of Ref.~\cite{vedral1}. Considering a two-level atom dipole interacting with a quantized mode of a high-$Q$ cavity. The atom-field interaction in the dipole approximation takes the form ($\hbar=1$) $V=g\sqrt{2}\left(\hat{a}^\dagger+\hat{a}\right)\hat{\sigma}_x$. Here, the effective coupling $g$ has been taken real, $\hat{a}^\dagger$ ($\hat{a}$) is the photonic creation (annihilation) operator for the field, and $\hat{\sigma}_x$ is the regular $x$-component of the Pauli matrices which acts on the two internal atomic states $|1\rangle$ and $|2\rangle$. Together with the free field and atom energies we have the Rabi Hamiltonian~\cite{jonas1} 
\begin{equation}\label{rabi}
\hat{H}_R=\omega\hat{a}^\dagger\hat{a}+\frac{\nu}{2}\hat{\sigma}_z+g\sqrt{2}\left(\hat{a}^\dagger+\hat{a}\right)\hat{\sigma}_x,
\end{equation}
where $\omega$ is the mode frequency, $\nu$ the atomic transition frequency, and $\hat{\sigma}_z$ is the Pauli $z$-matrix. In an interaction picture with respect to $\omega\left(\hat{a}^\dagger\hat{a}+\frac{\hat{\sigma}_z}{2}\right)$, we derive the JC model after imposing the RWA, i.e. neglecting counter rotating terms,
\begin{equation}\label{jc}
\hat{H}_{JC}=\frac{\Delta}{2}\hat{\sigma}_z+g\sqrt{2}\left(\hat{a}^\dagger\hat{\sigma}^-+\hat{\sigma}^+\hat{a}\right),
\end{equation}
with the atom-field detuning $\Delta=\nu-\omega$, and $\hat{\sigma}^\pm=(\hat{\sigma}_x\pm i\hat{\sigma}_y)/\sqrt{2}$ the lowering/raising atomic operators. The unitary transformation $\hat{U}(\varphi)=\exp\left[-i\varphi\hat{a}^\dagger\hat{a}\right]$ applied to the JC Hamiltonian gives
\begin{equation}\label{jc2}
\hat{H}_{JC}'=\frac{\Delta}{2}\hat{\sigma}_z+g\sqrt{2}\left(\hat{a}^\dagger e^{-i\varphi}\hat{\sigma}^-+\hat{\sigma}^+\hat{a}e^{i\varphi}\right).
\end{equation}  
The interpretation of the operator $\hat{U}(\varphi)$ is that it phase-shifts the field by $\varphi$. For each excitation quantum number $n$, the corresponding eigenstates $|\Phi_n^\pm(\varphi)\rangle$ traverse a loop $C$ on the associated Bloch sphere when $\varphi$ is varied slowly from 0 to $2\pi$. The accumulated Berry phases~\cite{berry,berrybook} 
\begin{equation}\label{berry}
\gamma_\pm=i\int_C d\varphi\,\langle\Phi_n^\pm(\varphi)|\frac{d}{d\varphi}|\Phi_n^\pm(\varphi)\rangle
\end{equation}
become~\cite{vedral1}
\begin{equation}\label{berry1}
\gamma_\pm=\pm\pi\left(1-\frac{\Delta/2}{\sqrt{\frac{\Delta^2}{4}+2g^2(n+1)}}\right)+\zeta_\pm(n),
\end{equation}
where $\zeta_\pm(n)$ is an integer times $2\pi$. In the case of $n=0$, it is seen that the phases $\gamma_\pm$ are non-trivial (i.e. not a multiple of $2\pi$), which is the main result of Ref.~\cite{vedral1}. We should make clear that the quantum number $n$ does not label the number of photons, e.g. the eigenstates $|\Phi_1^\pm\rangle$ both contain non-vacuum components of the field. However, what is shown in~\cite{vedral1} is that an atom initially in the excited state $|2\rangle$ and the field in vacuum still acquires a non-vanishing Berry phase according to the JC model. Moreover, using the fact that the atomic ground state $|1\rangle$ in the same vacuum mode does not pick up any Berry phase it is suggested how to achieve an interference experiment capable of measuring this vacuum induced Berry phase.  

Before analyzing the situation without the RWA, we present a semi-classical alternative demonstration of the Berry phase by defining a set of energy surfaces. To this end we give the JC Hamiltonian (\ref{jc}) in a quadrature representation~\cite{jonas1} 
\begin{equation}
\begin{array}{ccc}
\hat{a}=\frac{1}{\sqrt{2}}\left(\hat{x}-i\hat{p}\right),& &  
\hat{a}^\dagger=\frac{1}{\sqrt{2}}\left(\hat{x}+i\hat{p}\right),
\end{array}
\end{equation}
such that $\hat{x}$ and $\hat{p}$ obey the standard canonical commutation relations $[\hat{x},\hat{p}]=i$. Thus, we have
\begin{equation}\label{jcham2}
\hat{H}_{JC}=\frac{\Delta}{2}\hat{\sigma}_z+g\left(\hat{x}\hat{\sigma}_x+\hat{p}\hat{\sigma}_y\right).
\end{equation}
Interestingly, the JC interaction can be pictured as a special sort of spin-orbit coupling containing both the ``position" $\hat{x}$ and the ``momentum" $\hat{p}$. The {\it adiabatic energy potentials} are obtained as the eigenvalues of $H_{JC}$, treating $\hat{x}$ and $\hat{p}$ as $c$-numbers. This is the so called Born-Oppenheimer approximation of molecular physics~\cite{bear}. The resulting adiabatic potentials 
\begin{equation}\label{aps1}
E_\pm^{(JC)}=\pm\sqrt{\frac{\Delta^2}{4}+g^2\left(x^2+p^2\right)}=\pm\sqrt{\frac{\Delta^2}{4}+g^2r^2}
\end{equation}
can be seen as semi-classical energy surfaces such that $E_\pm^{(JC)}=\mathrm{const.}$ defines the semi-classical phase space trajectories. In the second equality of (\ref{aps1}) we introduced polar coordinates $re^{\pm i\phi}=x\pm ip$. Note that the two phases $\varphi$ and $\phi$ are highly related in that they both ``rotate" the coordinates $x$ and $p$. Note however, while $\varphi$ is an external control parameter, $\phi$ is a dynamical variable which for a localized (non-zero amplitude) field state, e.g. a coherent state, approximately change linearly in time for the JC model. 

\begin{figure}[h]
\centerline{\includegraphics[width=5cm]{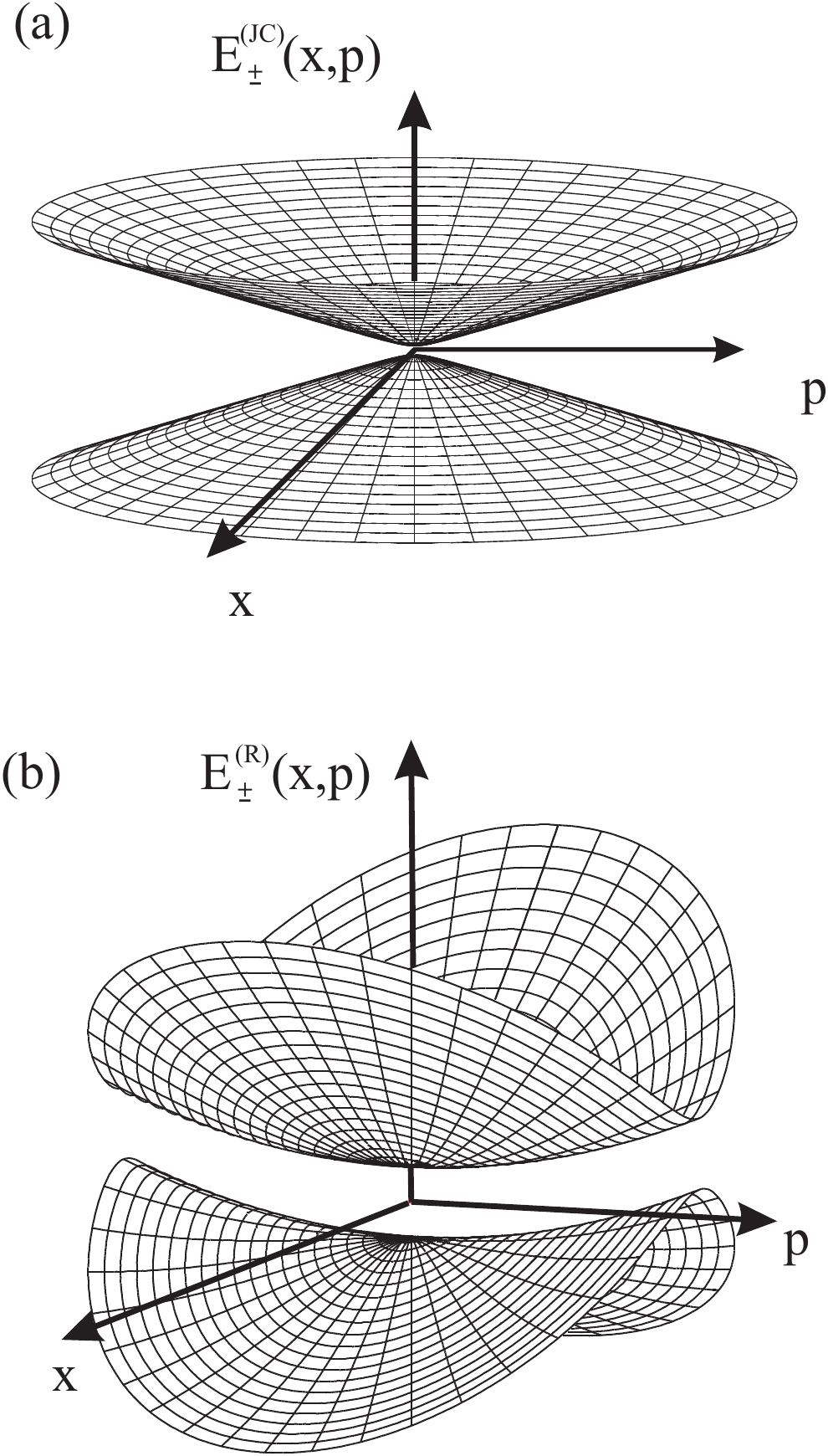}}
\caption{Semi-classical energy surfaces of the JC model (a) and the Rabi model (b). The CI in (a) indicates a non-vanishing Berry phase in this model, while absence of CI's in (b) signals that there are no corresponding Berry phases in the Rabi model.} \label{fig1}
\end{figure}

In Fig.~\ref{fig1} (a) we show the semi-classical energy surfaces $E_\pm^{(JC)}(x,p)$. The phase space trajectories are given for fixed $r$. At $r=0$ the two surfaces come the closest together (separated by $\Delta$). This point, $(x,p)=(0,0)$, characterizes a CI~\cite{bear}. The physics of CI's has been thoroughly studied in molecular and chemical physics for half a century~\cite{berrybook,bear,jt}, and it is known that adiabatically encircling a CI of the above {\it Jahn-Teller type}~\cite{jt} (i.e. letting $\phi$ change from 0 to $2\pi$) at a distance $r=R$ gives a Berry phase~\cite{koppel}
\begin{equation}\label{ciberry}
\gamma_\pm^{(CI)}=\pm\pi\left(1-\frac{\Delta/2}{\sqrt{\frac{\Delta^2}{4}+g^2R^2}}\right).
\end{equation}
We can identify the phase-space radius $R^2/2=n+1/2$. The resulting ``$1/2$" difference between Eqs.~(\ref{berry1}) and (\ref{ciberry}) (inserting $R^2/2=n+1/2$) derives from the fact that we performed an adiabatic diagonalization (imposing the Born-Oppenheimer approximation) instead of an exact diagonalization as is the case for (\ref{berry1}). The Born-Oppenheimer approximation breaks down in the vicinity of the CI~\cite{bear}, and our semi-classical results are therefore more accurate for large distances $R$ for which the ``$1/2$" term can be neglected and the two approaches agree. We may further note that CI's frequently appear in condensed matter physics, but is there more commonly referred to as {\it Dirac points}~\cite{topins}.  

We now turn to the more complete Rabi model~(\ref{rabi}), i.e. the JC model without the RWA,
\begin{equation}\label{rabi2}
\hat{H}_R=\omega\left(\frac{\hat{p}^2}{2}+\frac{\hat{x}^2}{2}\right)+\frac{\nu}{2}\hat{\sigma}_z+2g\hat{x}\hat{\sigma}_x.
\end{equation}
Applying the same phase shift transformation as for the JC Hamiltonian, $\hat{H}_R'=\hat{U}(\varphi)\hat{H}_R\hat{U}^{-1}(\varphi)$, the Rabi Hamiltonian becomes
\begin{equation}\label{rabi3}
\hat{H}_R'=\omega\left(\frac{\hat{p}^2}{2}+\frac{\hat{x}^2}{2}\right)+\frac{\nu}{2}\hat{\sigma}_z+2g\left(\cos\varphi\hat{x}-\sin\varphi\hat{p}\right)\hat{\sigma}_x.
\end{equation}
We have that when $\varphi$ is varied adiabatically between 0 and $2\pi$, a localized state in phase space will encircle the origin. Written in the traditional form of a spin-1/2 particle in an effective magnetic field, $\hat{H}_R'={\bf B}\bf{\cdot}\hat{\bf{\sigma}}$ (${\bf B}=(B_x,B_y,B_z)$ and $\hat{{\bf \sigma}}=(\hat{\sigma}_x,\hat{\sigma}_y,\hat{\sigma}_z)$), it follows, since $B_y=0$, that the $\bf{B}$-field of the Rabi model only moves within the $y=0$ plane. In order to cover a non-zero solid angle $\Omega(C)$, which would result in a non-trivial Berry phase, the magnetic field must contain a $y$-component. It is exactly such necessary term that appears when the RWA has been applied, as is evident from Eq.~(\ref{jcham2}). This absence of a Berry phase is also readily seen from $\hat{H}_R'$ of Eq.~(\ref{rabi3}). The semi-classical Hamiltonian is purely real and its adiabatic eigenstates
can be written $|\Theta_n^+(\phi)\rangle=\cos\left(\frac{\theta}{2}\right)|2,n-1\rangle+\sin\left(\frac{\theta}{2}\right)|1,n\rangle$ and $|\Theta_n^-(\phi)\rangle=\sin\left(\frac{\theta}{2}\right)|2,n-1\rangle-\cos\left(\frac{\theta}{2}\right)|1,n\rangle$, with the $x$- and $p$-dependent phase $\tan(\theta)=2(\cos\phi x-\sin\phi p)/\nu$. In this gauge, the adiabatic eigenstates are purely real and the Berry phase according to Eq.~(\ref{berry}) must be strictly zero. 

Furthermore, performing the same kind of semi-classical investigation for the Rabi model~(\ref{rabi2}) as we did for the JC model, it is found that the adiabatic potentials
\begin{equation}
E_\pm^{(R)}=\displaystyle{\omega\left(\frac{p^2}{2}+\frac{x^2}{2}\right)\pm\sqrt{\frac{\nu^2}{4}+4g^2x^2}}
\end{equation}
do not supply any CI's. As is demonstrated in Fig.~\ref{fig1} (b), the two surfaces do not intersect in a single point but along a line determined by $x=0$, and consequently the Berry phase should vanish.

In a strict sense, the above results rely on semi-classical arguments, and in order to verify the conclusions in a more general setting we numerically diagonalize the Hamiltonian~(\ref{rabi3}) for $0\leq\phi<2\pi$ and use the obtained eigenstates to calculate the corresponding Berry phases according to Eq.~(\ref{berry}). For various low lying eigenstates, the Berry phases are found to be zero regardless of parameter choices, including typical experimental parameters ranging from microwave and optical cavities~\cite{micro,optical} to the strong coupling regime of circuit QED~\cite{circuit}.    

Since Ref.~\cite{vedral1}, there have been a couple of proposals for observing the vacuum induced Berry phases in cavity QED systems~\cite{ext1,ext2}. These utilize classical driving fields and Raman coupled schemes. In a RWA, the phase of the classical fields can serve as the phase shift $\varphi$ needed to form the loop C in configuration space. Let us consider the simpler of these schemes, Ref.~\cite{ext1}, and argue that the resulting Berry phase is yet again an outcome of the RWA. The two lower atomic states of a $\Lambda$-atom are coupled to the excited state via a quantized cavity mode and a classical field respectively. If the driving fields are far detuned from the atomic transitions, an adiabatic elimination of the excited state render an effective two-level model and within the RWA the phase of the classical field represents $\varphi$ in Eq.~(\ref{jc2}), see~\cite{ext1} for details. Now, without any RWA, in the bare basis of the three atomic states the atom-field interaction in the dipole approximation takes the general form
\begin{equation}
V=\left[\begin{array}{ccc}
E_1 & 0 & \kappa\cos(\vartheta t+\varphi)\\
0 & E_2 & 2g\hat{x}\\
\kappa\cos(\vartheta t+\varphi) & 2g\hat{x} & E_3        
\end{array}\right],
\end{equation}
where $E_i$ is the bare energy of the $i$'th atomic state, $\kappa$ the effective field coupling between the atom and the classical field, and $\vartheta$ the driving frequency of the classical field. For simplicity, let us assume $E_1=E_2$ for which the time-dependent adiabatic energy potentials become
\begin{equation}
\begin{array}{l}
\displaystyle{E_\pm=\omega\left(\frac{p^2}{2}+\frac{x^2}{2}\right)+\frac{1}{2}\left(\delta\pm\sqrt{\delta^2+G^2}\right)},\\ \\
\displaystyle{E_0=\omega\left(\frac{p^2}{2}+\frac{x^2}{2}\right)},
\end{array}
\end{equation}
with $\delta=E_3-E_1$ and $G^2=\kappa^2\cos^2(\vartheta t+\varphi)+4g^2x^2$. As in the case of the Rabi model, the three surfaces do not possess a point of intersection, e.g. CI,  and there cannot exist a non-zero Berry phase when $\varphi$ is varied.

This far we have seen how application of the RWA may lead to incorrect conclusions in terms of Berry phases. It does not, however, exclude interesting geometric phase effects in cavity QED~\cite{jonas2,jonas3,dicke1,dicke2}. In Ref.~\cite{jonas2}, a bi-modal cavity QED system was analyzed without the RWA, and it was found that under certain mode polarizations the model is identical to the $E\times\varepsilon$ Jahn-Teller one of molecular physics. The corresponding Berry phase in this model was shown to greatly affect the field properties of the two modes~\cite{jonas2,jonas3}. The question regarding vacuum induced Berry phases was not, on the other hand, addressed in~\cite{jonas2}. The corresponding Berry phases when encircling the CI at a distance $r=R$ is~\cite{jonas2}
\begin{equation}\label{jtberry}
\gamma_\pm^{(JT)}(R)=\pm\pi\left(1-\frac{\nu}{\sqrt{\nu^2+4g^2R^2}}\right).
\end{equation}
The parameter $R$ is indirectly related to the two field amplitudes, and for the two modes to be in vacuum it seems plausible to identify the vacuum induced Berry phase by letting $R\rightarrow0$. This, since $\gamma_\pm^{(JT)}(0)=0$ as long as $\nu\neq0$, indicates that there are no vacuum induced Berry phases in this model. Reference~\cite{dicke1} considers the ground state of the Dicke model without the RWA and with atomic dipole-dipole interaction. They find a Berry phase strictly zero in the {\it normal phase} (that is with the cavity mode in vacuum), and a non-zero phase in the {\it superradiant phase} where the field is no longer in the vacuum. Finally we have another study of the ground state Dicke model without the RWA~\cite{dicke2}. The generation of the Berry phase differs in this work compared to~\cite{vedral1}; the Hamiltonian is unitary transformed with an atom rotation $\hat{U}_{at}(\varphi)=\exp\left[-i\varphi\hat{S}_z/2\right]$ rather than a field rotation $\hat{U}(\varphi)$ utilized above, i.e. the Bloch vector is directly rotated. Yet again it is found that the resulting Berry phase vanishes in the normal phase. Thus, neither of these three works support a vacuum induced Berry phase.       

The remarkable aspect of our discoveries is that they are independent of system parameters. In the majority of experiments, the ratio $g/\omega\ll1$ which implies that the counter rotating terms of the Hamiltonian is assumed to be negligible, i.e. justifying the application of the RWA. In the present paper we give probably the first example where this assumption is clearly false. We conclude that this derives from the fact that we consider adiabatic evolution. Similarly, the Berry connection $\langle\Phi(\varphi)|\frac{d}{d\varphi}|\Phi(\varphi)\rangle$ vanishes in the adiabatic limit provided non-degenerate states, but still closed line integrals of it may render a non-vanishing Berry phase. Consequently, despite the fact that the effects deriving from counter rotating terms may be negligible at finite time scales, they indeed become important for adiabatic processes.  

In conclusion, we have demonstrated that imposing the RWA in certain cavity QED systems can impart incorrect results regardless of system parameters. Focus was on vacuum induced Berry phases predicted in numerous works, and it was shown that they exactly disappear when the RWA is not implemented. We learn from this, that utilizing RWA's (and also adiabatic elimination schemes) must be performed with great caution when considering Berry phases, or more general for adiabatic scenarios. This said, our results does not exclude existence of vacuum induced Berry phases, despite the fact that they seem to vanish also in the settings of Refs.~\cite{jonas2,dicke1,dicke2} of which non employ any RWA, but the topic is surely more subtle than first thought. It should be pointed out that our analysis has been carried out for the JC model, which in itself is an approximation of more realistic situations where many degrees-of-freedom has been neglected. It might be that the inclusion of additional cavity modes or atomic electronic levels lead to qualitative changes in our conclusions. Furthermore, since our findings seem to derive from the adiabatic assumption, they do not forbid non-adiabatic vacuum induced geometrical phases also for idealized models as the JC one. Thus, it would be especially interesting to analyze the Aharonov-Anandan geometric phase~\cite{berryext} in the Rabi model. 

{\it Note added}. During the reviewing process of this manuscript, I became aware of the related works~\cite{solinas}, which show how application of the {\it secular approximation} for master equations may render inconsistent results in the adiabatic limit.

\begin{acknowledgements}
I acknowledge support from the Swedish research council (VR), Kungl. Vetenskapsakademien (KVA), and Deutscher Akademischer Austausch Dienst (DAAD).
\end{acknowledgements}

\end{document}